\documentclass[apj]{emulateapj}

\newcommand{\msun}{$M_{\odot}$}

\newcommand{\rxjfull}{RX~J0152.7-1357}
\newcommand{\rxj}{RX~J0152-13} 
\newcommand{\sfr}{$M_{\odot}$~yr$^{-1}$}
\newcommand{\ssfr}{yr$^{-1}$}
\newcommand{\galmpc}{Mpc$^{-2}$}
\newcommand{\pmpc}{Mpc$^{-1}$}
\newcommand{\kmps}{km~s$^{-1}$}

\newcommand{\uJy}{$\mu$Jy}

\newcommand{\Rfil}{$R$}
\newcommand{\zfil}{$z$}

\newcommand{\zwindow}{$0.80<z<0.87$}
\newcommand{\mcut}{10.3}
\newcommand{\ntot}{1234}   
\newcommand{\nmcut}{330}  
\newcommand{\mcutrange}{$M > 2 \times 10^{10}$~\msun}
\newcommand{\mcutrangelog}{$\log M/M_{\odot} >~$\mcut}

\begin{document}

\title{The Dependence of Star Formation Rates on Stellar Mass and Environment at $\lowercase{z} \sim 0.8$ \altaffilmark{1,2,3,4}}

\author{Shannon G. Patel \altaffilmark{5},
Bradford P. Holden\altaffilmark{5},
Daniel D. Kelson\altaffilmark{6},
Garth D. Illingworth\altaffilmark{5},
Marijn Franx\altaffilmark{7},
}

\altaffiltext{1}{This paper includes data gathered with the 6.5~meter Magellan Telescopes located at Las Campanas Observatory, Chile.}
\altaffiltext{2}{This work is based on observations made with the Spitzer Space Telescope, which is operated by the Jet Propulsion Laboratory, California Institute of Technology under a contract with NASA. Support for this work was provided by NASA through an award issued by JPL/Caltech.}
\altaffiltext{3}{Based in part on data collected at Subaru Telescope, which is operated by the National Astronomical Observatory of Japan.}
\altaffiltext{4}{Some of the data presented herein were obtained at the W. M. Keck Observatory, which is operated as a scientific partnership among the California Institute of Technology, the University of California and the National Aeronautics and Space Administration. The Observatory was made possible by the generous financial support of the W.M. Keck Foundation.}
\altaffiltext{5}{UCO/Lick Observatory, University of California, Santa Cruz, CA 95064; patel@ucolick.org}
\altaffiltext{6}{Observatories of the Carnegie Institution of Washington, Pasadena, CA 91101}
\altaffiltext{7}{Leiden Observatory, Leiden University, P.O. Box 9513, NL-2300 AA Leiden, Netherlands}

\begin{abstract}
We examine the star formation rates (SFRs) of galaxies in a redshift slice encompassing the $z = 0.834$ cluster \rxjfull.  We used a low-dispersion prism in the Inamori Magellan Areal Camera and Spectrograph (IMACS) to identify galaxies with $z_{\rm AB} < 23.3$~mag in diverse environments around the cluster out to projected distances of $\sim 8$~Mpc from the cluster center.  We utilize a mass-limited sample (\mcutrange) of \nmcut\ galaxies that were imaged by Spitzer MIPS at 24~\micron\ to derive SFRs and study the dependence of specific SFR (SSFR) on stellar mass and environment.  We find that the SFR and SSFR show a strong decrease with increasing local density, similar to the relation at $z \sim 0$.  Our result contrasts with other work at $z \sim 1$ that find the SFR-density trend to reverse for luminosity-limited samples. These other results appear to be driven by star-formation in lower mass systems ($M \sim 10^{10}$~\msun).  Our results imply that the processes that shut down star-formation are present in groups and other dense regions in the field.  Our data also suggest that the lower SFRs of galaxies in higher density environments may reflect a change in the ratio of star-forming to non-star-forming galaxies, rather than a change in SFRs.  As a consequence, the SFRs of star-forming galaxies, in environments ranging from small groups to clusters, appear to be similar and largely unaffected by the local processes that truncate star-formation at $z \sim 0.8$.
\end{abstract}


\section{Introduction}

Pioneering studies of the impact of environment on galaxy properties have found higher density environments in the local universe to be dominated by elliptical and S0 galaxies \citep{dressler1980}.  Because of the strong correlation of Hubble type with star formation rate \citep[SFR,][]{kennicutt1998}, a correlation between SFR and local galaxy density is expected and has been seen in more recent studies of the local universe \citep{gomez2003,balogh2004}.  \citet{kauffmann2004} examined the SDSS sample of \citet{brinchmann2004} and found a strong dependence of SFR on local galaxy density and stellar mass at $z \sim 0$.  At a fixed stellar mass, they found the specific SFRs (SSFRs), and therefore SFRs, of galaxies to be lower in higher density environments.

At $z \sim 1$, the morphology-density relation (MDR) follows a similar trend to that found in the local universe for mass-limited samples \citep{holden2007,vanderwel2007b}.  In contrast, recent analyses of the field at $z \sim 1$ suggest that the SFR-density trend was the {\em reverse} at earlier times, where galaxies at higher densities display {\em higher} SFRs than galaxies at lower densities \citep{elbaz2007,cooper2008}.

In this Letter, we explore the SFRs of galaxies in a redshift slice that includes the $z = 0.834$ galaxy cluster \rxjfull\ (hereafter \rxj).  Our wide-field spectroscopic survey extends to a projected distance of $\sim 8$~Mpc from the cluster center, resulting in a large number of galaxies that span a much broader range of environments at this epoch than is found in other work.  Our aim is to determine the form of the SSFR/SFR-density relation at $z \sim 0.8$ and compare to the results discussed above at these redshifts and at $z \sim 0$.

We assume a cosmology with $H_0 = 70$~\kmps~\pmpc, $\Omega_M = 0.3$, and $\Omega_{\Lambda} = 0.7$.  Stellar masses and SFRs are based on a Chabrier IMF \citep{chabrier2003}.

\section{Data and Analysis}
Galaxies with $z_{\rm AB} < 23.3$~mag from archival Subaru Suprime-Cam wide-field imaging of \rxj\ in $VRiz$ were targeted over a $29\arcmin \times 39\arcmin$ FOV with a low-dispersion prism (LDP) in the Inamori Magellan Areal Camera and Spectrograph \citep[IMACS,][]{dressler2006} on Magellan-Baade.  Redshifts were determined by fitting the LDP spectra and Suprime-Cam broadband photometry with a grid of \citet[BC03]{bc03} model templates as described in \citet{patel2009}.  The redshift extraction utilized a strong set of priors on the model galaxy templates, therefore we refrained from using any rest-frame quantities from the fitting in order to avoid any biases.  We found a total of \ntot\ galaxies in the cluster redshift interval of \zwindow.  The survey has a high level of completeness (median of $\sim 85\%$ for the sample discussed in this work) and is remarkably uniform for different galaxy types (e.g. red/blue, faint/bright) while varying spatially from $\sim 90\%$ in the cluster core to $\sim 75\%$ at projected clustercentric radii of $R \sim 6$~Mpc.  Where required, galaxies were given weights based on a completeness map that is a function of \zfil-band magnitude, \Rfil-\zfil\ color, and sky position.  When compared to redshifts derived from higher resolution spectra, the LDP redshifts display a $1\sigma$ biweight scatter of only $1.1\%$ in $(1+z)$ for galaxies in the redshift interval above.  The scatter is the same for fainter galaxies ($z_{AB} > 22$~mag), while red and blue galaxies have a scatter of $0.89\%$ and $1.1\%$, respectively.  Thus, the LDP redshift uncertainties lack any significant dependence on the magnitudes and colors probed here, demonstrating their superiority over photometric redshifts that are typically challenged by these systematics. 

We used Spitzer MIPS imaging at 24~\micron\ to calculate SFRs.  The MIPS imaging covers $732$ galaxies in a range of environments.  We used the technique employed in \citet{vanderwel2007} to derive SFRs.  At the MIPS $5\sigma$ detection limit of $125$~\uJy, we measure a SFR of $\sim 12$~\sfr\ for galaxies at $z \sim 0.8$.  While SFRs derived from the MIPS data represent the obscured component of star-formation (SF), several authors have found this component to dominate the total SFR, especially for galaxies in the mass range discussed in this paper \citep{bell2005,elbaz2007}.  Meanwhile, more recent findings suggest SFRs inferred from the mid-IR may reflect SF activity over longer timescales \citep[1$-$2~Gyr,][Kelson \& Holden 2009 in prep]{salim2009}.  Finally, AGN do not appear to contribute significantly to the overall mid-IR luminosity density at these redshifts \citep{bell2005} or to the mid-IR luminosity of individual galaxies \citep{salim2009}.

Stellar masses were computed by fitting our prism spectra and $VRiz$ photometry with a suite of BC03 $\tau$-models at the derived prism redshift.  At the redshift of \rxj\ ($z = 0.834$), our sample reaches a limiting mass of \mcutrange\ (\mcutrangelog) on the red-sequence.  Above this mass limit, our sample consists of \nmcut\ galaxies with MIPS imaging.

We compute the local projected galaxy density for each galaxy by measuring the distance to the 7th nearest neighbor with \mcutrange\ and relative rest-frame velocity $|\Delta v| < 6000$~\kmps\ (i.e. 2\% of $(1+z)$).

\section{MIPS Stacking}

\begin{figure}
\epsscale{1.2}
\plotone{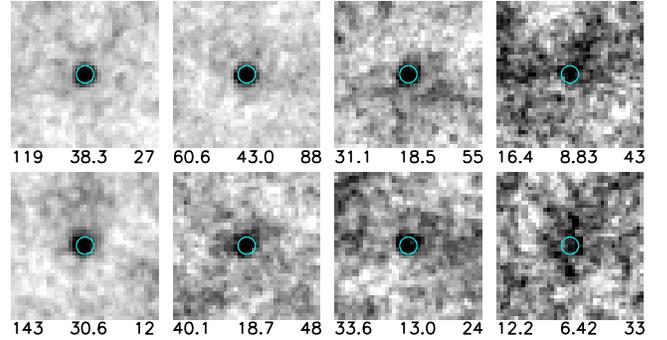}
\caption{Median stacked MIPS images ($51$\arcsec on a side) for two bins of mass (increasing towards bottom) and four bins of density (increasing towards right): this layout mimics the position of the data points in Figure~\ref{stack_mips_density}.  The numbers at the bottom of each image indicate, from left to right: (1) total flux (\uJy), corrected from the $D = 6\arcsec$ apertures shown, (2) S/N in the aperture based on the rms of the background, and (3) number of galaxies in each stack.  The median galaxy is detected (S/N~$>5$) in each bin.  The corresponding SFR and its uncertainty for a given mass-density bin were derived from a bootstrap analysis and are shown in Figure~\ref{stack_mips_density}.} \label{mipsstack_ps}
\end{figure}

We performed a stacking analysis of our MIPS imaging to determine the median SFRs for sub-samples divided by stellar mass and environment.  Most galaxies in higher density environments were not individually detected.  Therefore, a stacking analysis for galaxies in all environments allowed us to determine the median SFRs uniformly.  All galaxies were included in the stacks, regardless of being detected at 24~\micron.  We used a bootstrap approach to determine the median SFR and its uncertainty for a stack of images.  As a check on this technique, we compared the median flux computed from a stack of detected sources to the median value of the individual fluxes, and found them to agree to within $5\%$.  We used the median of the stack, as opposed to the mean, because it was less susceptible to being influenced by bright contaminating sources.

Figure~\ref{mipsstack_ps} shows the median image for different mass-density bins.  We find a detection for the median stacked galaxy in each bin.  Galaxies were stacked in 2 bins of mass above our mass limit, $10.3 < \log M/M_{\odot} < 10.8$ and $\log M/M_{\odot} > 10.8$, and 4 bins of local density that were divided at $\Sigma \sim 3$, 10, and 32~\galmpc, with the density range of the extreme bins unbounded.  We used wide density bins given the large scatter between dark matter halo mass and local projected galaxy density \citep[see e.g.][Fig.~16]{kauffmann2004}.

\section{Specific Star Formation Rates, Mass, and Environment}

\begin{figure}
\epsscale{1.2}
\plotone{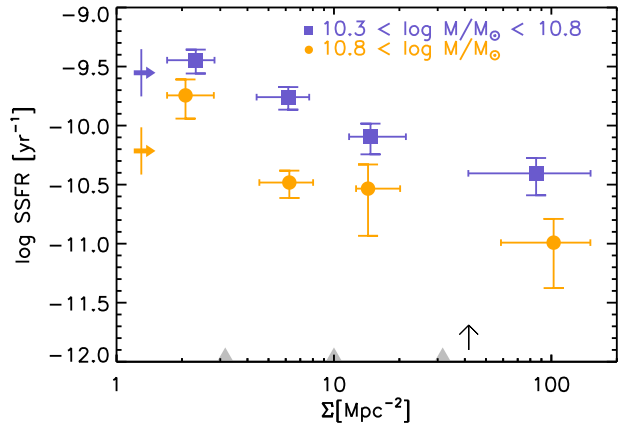}
\caption{Median SSFR vs. local density at $z \sim 0.8$ for galaxies in two mass bins.  The location of the data points on the x-axis reflects the median local density in a particular bin ({\em gray triangles} denote boundaries), while the horizontal error bars represent the $25-75$th percentile of density values in the bin.  The arrows on the left indicate the median SSFRs for the field from \citet{damen2009} for two similar mass bins: they lie between our two lowest density bins, indicating that our data sample a range of field environments.  The $0.2$~dex error assigned to their data points account for systematics in the different SFR and stellar mass derivations.  The {\em arrow} at $\Sigma \sim 40$~\galmpc\ corresponds to the median density at the projected virial radius of the cluster, but we note that the highest density bin includes contributions from both the cluster and groups.  At a fixed mass, the SSFR and SFR decline in higher density environments, and this is true even at low densities where the presence of the cluster is inconsequential.} \label{stack_mips_density}
\end{figure}

\begin{figure}
\epsscale{1.2}
\plotone{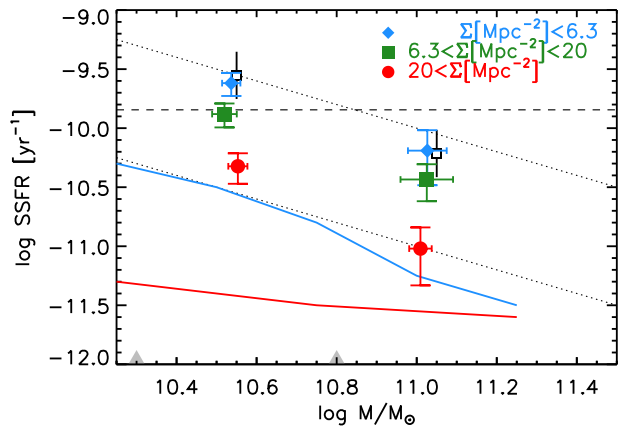}
\caption{Median SSFR vs. stellar mass at $z \sim 0.8$ for galaxies in different environments.  The location of the data points on the x-axis reflects the median mass in a particular bin, while the horizontal error bars represent its uncertainty.  The {\em dotted lines} indicate constant SFRs of 1 and 10~\sfr.  The {\em dashed horizontal line} represents the SSFR at which stellar mass doubles by at $z=0$ if the SFR remains constant.  The {\em black} data points are the \citet{damen2009} field values: note the good agreement with our field values.  The {\em red} and {\em blue colored lines} are the median SSFR values from \citet{kauffmann2004} for galaxies at $z \sim 0$ at their lowest and highest densities, respectively.  At $z \sim 0.8$, SSFR depends on mass and environment, with galaxies at high masses and high densities forming stars at lower rates, much like at $z \sim 0$.} \label{stack_mips}
\end{figure}

Figure~\ref{stack_mips_density} shows the median SSFR vs. local density for galaxies in two mass bins.  The median mass of galaxies in each mass-density bin was used to compute SFR/M (i.e. SSFR).  The error bars for the SSFRs reflect the uncertainty on the SFRs from the bootstrap analysis and the uncertainty in the median mass of each bin.  There is a clear correlation between SSFR and density, with galaxies at higher densities having lower SSFRs.  This is true for galaxies in both mass bins.  Furthermore, at fixed density the lower mass galaxies have higher SSFRs.  Note that because the SSFR-density trends are shown for a given mass bin, it implies that the SFR also declines across the entire density range.  It also implies that the overall decline in SSFR with density is not due to any change in the mass function between the low-density field and the high-density cluster environment since the SSFR declines with density at a fixed mass.

Figure~\ref{stack_mips} emphasizes this point.  The figure shows the median SSFR vs. stellar mass for galaxies in three density bins.  The density bins have been recast from Figure~\ref{stack_mips_density} to roughly reflect the division between the cluster, groups, and field.  At a fixed mass, galaxies in the highest density bin have SSFRs that are $\sim 6$ times lower than galaxies in the low-density bin.  Note that the median $z \sim 0.8$ field data from \citet{damen2009} agree with the median SSFRs of galaxies in our lowest density bin, indicating that this bin is representative of the general field.  Also note that the \citet{damen2009} points roughly lie between the SSFRs of the two lowest density bins in Figure~\ref{stack_mips_density}, implying that we in fact sample a range of field environments.

Figure~\ref{stack_mips} also shows the median SSFR trends for galaxies in the lowest ($blue$) and highest ($red$) density environments at $z \sim 0$ from \citet{kauffmann2004}.  These $z \sim 0$ total SFRs were derived from extinction corrected H$\alpha$ luminosities.  In our $z \sim 0.8$ sample the MIPS derived SFRs dominate the total, allowing us to compare our SSFRs to those from \citet{kauffmann2004}.  Although our density estimators were not computed in a similar manner, by comparing galaxies at the lowest and highest densities in both samples we can compare and contrast galaxies at the two extremes at both epochs.

Above the mass completeness limit, galaxies at all densities in our $z \sim 0.8$ sample are forming stars at a higher rate relative to galaxies with the same stellar mass at $z \sim 0$.  One should note however, that the bulk of SF has already taken place for most of these galaxies, even by $z \sim 0.8$, since most lie below the dashed line in Figure~\ref{stack_mips} that indicates the SSFR at which stellar mass doubles by at $z = 0$ if the SFR remains constant.

At $z \sim 0$, \citet{kauffmann2004} found a factor of $\sim 7$ spread in SFRs at $\log M/M_{\odot} \sim 10.6$ between their lowest and highest density bins, with galaxies at high densities having lower SFRs.  At $z \sim 0.8$, we found a similar result.  For galaxies with mass $\log M/M_{\odot} \sim 11$, the spread in SFRs at $z \sim 0$ is smaller than the factor of $\sim 7$ we find.  This is likely because \citet{kauffmann2004} assigned a limiting SSFR of $\sim 10^{-11.6}$~\ssfr\ to galaxies with non-detectable SF.  The median SFRs have not been corrected for contamination from galaxies in lower density bins, but we note that such a correction would lead to an even stronger decline in SSFR with increasing density.

\section{Discussion}

Recent work at $z \sim 1$ indicates that for galaxies in the mass range studied here, several relations follow trends similar to those found at $z \sim 0$.  For example, \citet{holden2007} and \citet{vanderwel2007b} found a strong MDR for mass-limited samples in both the field and clusters at $z \sim 1$.  Thus, it should not be surprising that we see a similar trend for the SSFR/SFR-density relation at $z \sim 0.8$ as we see at $z \sim 0$ (although with a different normalization).  Interestingly, some recent studies of galaxies at low-redshift find evidence for enhanced levels of dust-obscured SF at densities slightly above typical field densities, although they also generally find an overall trend of decreasing SF activity at higher densities \citep{gallazzi2009,wolf2009,haines2009b}.

Our result of a declining SFR in higher density environments appears universal at $z \sim 0.8$ and not confined to a cluster environment, much like the MDR and SFR-density relations in the local universe.  As seen in Figure~\ref{stack_mips_density}, cluster galaxies dominate the highest density bin, but much of the decrease in SSFR occurs in lower density field bins.  In addition, after removing galaxies within $\sim 2.5$ times the virial radius of the cluster, we continue to find the SSFR to decrease with density, including in the remaining high density regions represented by several groups at projected distances of $\sim 3-5$~Mpc from the cluster.  This decrease in SSFR occurs over a similar range of densities in which \citet{patel2009} found an increase in the red galaxy fraction, possibly linking the end of SF and buildup of red-sequence galaxies in environments that reach into the dense regions of the field.

In contrast to our work, \citet{elbaz2007} and \citet{cooper2008} found very different results for the SFR-density relation.  At $0.8 < z < 1.2$, \citet{elbaz2007} found a factor of $\sim 6$ spread in SFRs.  However, they found galaxies at higher densities to have {\em higher} mean SFRs up to a critical density, above which the SFR declined.  Much of the reversed SFR-density trend in \citet{elbaz2007} is driven by a peak in the SFR in a narrow projected density range of $\sim 0.1$~dex ($3 < \Sigma~(\rm Mpc^{-2}) < 4$).  Likewise, \citet{cooper2008} also found the SFRs of galaxies to increase in higher density environments at $0.75 < z < 1.05$, although their observed spread in SFRs was less than a factor of $\sim 1.5$.  While neither of these two surveys contain a cluster, both sample group environments similar to the groups around \rxj\ that have velocity dispersions \citep[$\sim 400$~\kmps,][]{tanaka2006} that are typical of groups found in the field \citep{gerke2007}.  However, the reversal in the SFR-density relation found by \citet{elbaz2007} and \citet{cooper2008} does not extend into the highest densities found in their group environments.

Sample selection plays an important role in interpreting the different results.  While we use a mass-limited sample, \citet{elbaz2007} and \citet{cooper2008} use luminosity-limited samples that are biased to include low-mass blue star-forming galaxies and exclude the corresponding non-star-forming red galaxies.  When restricted to a mass-limited sample that was similar to the one in this work, \citet{elbaz2007} found the SSFR continued to increase from low-density to high-density but with marginal significance (see Fig.~20 in that work).  \citet{cooper2008} found the mean SFR to increase with density by $\sim 35\%$ while for the same sample the mean SSFR {\em decreases} by $\sim 35\%$ over the same density range, which implies an increase in the mean mass with density by a factor of $\sim 2$.  The inferred range of mean mass ($9.9 \la \log M/M_{\odot} \la 10.2$) from \citet{cooper2008} is below our mass threshold.  However, the small rise in SFRs with density seen by \citet{cooper2008} can be explained by the presence of relatively more high mass galaxies, which have higher SFRs \citep{elbaz2007,noeske2007b}, at higher densities in their sample.

Interestingly, we note that when selecting a luminosity-limited sample that was similar to either of these works, we found the SFR to continue to decrease at higher densities, but at a much shallower pace for lower mass galaxies.  This did not depend strongly on the sub-sample of galaxies that were used to compute local densities (i.e. luminosity vs. mass-limited).  We note that contaminants from low-density environments, which we found to have higher levels of SF, likely contribute to the shallower trend.

Recent observations of a ``star-forming sequence'' in the field at $z \sim 0$ \citep{salim2007} and up to redshifts of $z \sim 1$ \citep{noeske2007} suggest that galaxies populate a narrow distribution of SSFRs at a fixed stellar mass.  Below this sequence is a more extended distribution of galaxies with very low SSFRs that are effectively non-star-forming.  Here, we investigate whether the SSFR-density trend represents (1) a changing mix of galaxies that lie on or off of the star-forming sequence or (2) a change in the SFRs of star-forming galaxies in different environments.  Without additional constraints, our MIPS stacking analysis alone cannot distinguish between these two scenarios.

Using our previous results on the correlation between the red galaxy fraction and density \citep{patel2009}, we can speculate on which of these scenarios is more likely.  In \citet{patel2009} we found the fraction of blue galaxies increased by a factor of $\sim 10$ from $\sim 5\%$ in the highest density regions to $\sim 50\%$ at the lowest densities.  If the typical SFR of galaxies on the star-forming sequence remains constant and the ratio of blue-to-red galaxies reflects the ratio of galaxies that are forming stars or not then one expects the average SFR (and SSFR within a single mass bin) in the high density regions to be diminished by a factor of $\sim 10$ compared to the field, roughly what we see in the MIPS stacking analysis presented here.  The notion that galaxies are either star-forming or not is given additional credence by \citet{bell2005} and \citet{dressler2009b} who found that the dominant mode of SF at these epochs was in moderate starbursts when accounting for the duty cycle of such events.

In summary, we find that at $z \sim 0.8$, the SSFRs and SFRs of galaxies with \mcutrange\ decrease in higher density environments, and this is true over the entire range of environments studied.  This result follows the trend observed at $z \sim 0$ in which \citet{kauffmann2004} also found galaxies at higher densities to have lower SSFRs, but with SSFRs that were $\sim 10$ times higher at $z \sim 0.8$.  Our result differs from that of \citet{elbaz2007} and \citet{cooper2008} who found a reversal in the SFR-density trend at $z \sim 1$ over some portion of their density range, mostly driven by SF in lower mass galaxies present in their luminosity-limited samples.  In a forthcoming paper, we plan to utilize rest-frame UV data with SED fitting to determine total SFRs of individual galaxies at $z \sim 0.8$.  The distribution of SFRs in different environments will allow us to probe the processes that were responsible for shutting down SF at a time when the universe was half its current age.

\acknowledgments
This research was supported by NASA grant NAG5-7697 and Spitzer grants JPL 1277397 and JPL 1344481.  We thank the referee for helpful feedback in improving the quality of the manuscript.  We also thank Kai Noeske, Arjen van der Wel, and Sandy Faber for useful discussions.  Finally, we acknowledge Scott Burles for developing the LDP, and the PRIMUS collaboration for allowing us to investigate galaxies with their hardware.



\end{document}